\documentclass[amsmath, amssymb,10pt,aps,prb,onecolumn,notitlepage,showpacs,superscriptaddress]{revtex4-1}  
\usepackage{graphicx} 
\usepackage[bookmarks=false]{hyperref}
\usepackage{amsmath}
\usepackage{amssymb}
\usepackage{bm}
\usepackage{amsthm}
\usepackage{amsfonts}
\usepackage{latexsym}
\usepackage{wrapfig}
\usepackage[usenames,dvipsnames]{color}
\usepackage[protrusion=true,expansion=true,final]{microtype}
\usepackage{ifthen}
\usepackage{tikz}\usetikzlibrary{shapes,arrows,positioning}
\usepackage{color,soul}
\usepackage{csquotes}
\usepackage{booktabs}
\usepackage[normalem]{ulem}
\usepackage{chemformula}
\useunder{\uline}{\ul}{}
\usepackage{textcomp}
\usepackage[most]{tcolorbox}
\usepackage{xcolor}

\usepackage{amsmath,amssymb,mathtools}
\usepackage{bm}
\usepackage{siunitx}
\usepackage{graphicx}
\usepackage{booktabs}
\usepackage{physics}
\usepackage[most]{tcolorbox}
\usepackage{microtype}
\usepackage{cleveref}

\begin{document}

\title{Spatially dispersive photogalvanic effects as a probe of quantum geometric tensor}

\author{Yicong Chen}
\thanks{These authors contributed equally.}
\affiliation{Department of Physics, University of Pennsylvania, Philadelphia, Pennsylvania 19104, USA}

\author{Bumseop Kim}
\thanks{These authors contributed equally.}
\affiliation{Department of Chemistry, University of Pennsylvania, Philadelphia, Pennsylvania 19104, USA}

\author{Senin Ahammed Akkara Paramban}
\affiliation{Department of Materials Science and Engineering, University of Pennsylvania, Philadelphia, Pennsylvania 19104, USA}

\author{Zhurun Ji}
\affiliation{Department of Physics, Massachusetts Institute of Technology, Cambridge, Massachusetts 02139, USA}

\author{Utkarsh Khandelwal}
\affiliation{Department of Materials Science and Engineering, University of Pennsylvania, Philadelphia, Pennsylvania 19104, USA}

\author{Shupeng Xu}
\affiliation{Department of Materials Science and Engineering, University of Pennsylvania, Philadelphia, Pennsylvania 19104, USA}

\author{Jiachen Shi}
\affiliation{Department of Materials Science and Engineering, University of Pennsylvania, Philadelphia, Pennsylvania 19104, USA}

\author{Sergiy Krylyuk}
\affiliation{Material Measurement Laboratory, National Institute of Standards and Technology, Gaithersburg, 20899, Maryland, USA}

\author{Albert V. Davydov}
\affiliation{Material Measurement Laboratory, National Institute of Standards and Technology, Gaithersburg, 20899, Maryland, USA}

\author{Andrew M. Rappe}
\affiliation{Department of Chemistry, University of Pennsylvania, Philadelphia, Pennsylvania 19104, USA}

\author{Ritesh Agarwal}
\email{riteshag@seas.upenn.edu}
\affiliation{Department of Materials Science and Engineering, University of Pennsylvania, Philadelphia, Pennsylvania 19104, USA}

\date{\today}

\begin{abstract}
Berry curvature and quantum metric of Bloch states govern a wide range of phenomena, yet unified experimental access to both remains limited. Here we show that light spatial-dispersion alters the selection rules of second-order photocurrents and responses forbidden by parity become allowed in centrosymmetric crystals without any external fields, magnetic order, interfaces, strain, or engineered symmetry breaking. The resulting response is intrinsic and its weight is set by the quantum geometry of the Bloch states. Conventional photogalvanic effects are constrained by crystal symmetry and entangle the quantum metric with shift vector contributions. The spatially-dispersive response instead isolates the quantum metric and Berry curvature in distinct polarization channels. Implementing this approach in $1T^\prime$-MoTe$_2$ across its temperature-driven transition to $T_d$-Weyl phase, we resolve helicity-even and helicity-odd photocurrents corresponding to quantum metric and Berry curvature contributions, respectively, from the same device. The metric-dominated response persists across both phases and exhibits a robust spectral structure reproduced by first-principles calculations and linked to momentum-resolved quantum metric hotspots. In contrast, the curvature-driven channel emerges only when inversion symmetry is broken and shows strong sensitivity to carrier doping through competing momentum-space contributions. Our results establish photogalvanic effects with spatially varying optical fields as a general route to accessing quantum geometry in materials where photon energy and electronic filling probe different parts of the excitation manifold.

\end{abstract}

\maketitle

\section{Introduction}

Quantum geometry---encoded by the Berry curvature (imaginary part) and the quantum metric (real part) of the quantum geometric tensor (QGT)---governs a wide class of transport and optical effects in solids~\cite{berry1989quantum, provost1980riemannian, verma2026quantum}. Historically, experimental access has been dominated by the Berry curvature, which vanishes under combined inversion and time-reversal symmetry and can therefore be isolated by symmetry breaking. The quantum metric, by contrast, is symmetry-even and generically finite and therefore cannot be isolated by symmetry alone. It has nonetheless become central to flat-band superconductivity, exciton condensation, fractional Chern insulators, and moiré physics. 

Recent approaches have begun to probe the quantum metric. ARPES reconstructions from pseudospin textures are surface sensitive and require effective-Hamiltonian fits. Nonlinear transport is bulk sensitive but integrates over the Brillouin zone near the Fermi level and is convoluted with disorder scattering. More fundamentally, existing transport routes rely on an externally imposed symmetry-breaking element — a magnetic field, a magnetic order, or a specific point-group condition — to activate or isolate the response; they report the response of a perturbed system and restrict the accessible material space.

Nonlinear optical responses provide a versatile route to probe quantum geometry, as they directly involve interband coherences. In particular, second-order photocurrents such as the linear and circular photogalvanic effects (LPGE and CPGE) have been shown to encode information about the underlying quantum geometry of Bloch states~\cite{ahn2022riemannian,ma2021topology}. The antisymmetric part of the overall PGE response (CPGE) is closely related to the Berry curvature and has been used to probe topological and symmetry properties of a variety of materials, while the symmetric part of the response (LPGE) reflects geometric aspects of interband transitions and has been discussed in connection with the quantum metric~\cite{dong2026layer}. The symmetric dipole product encodes interband coherence and shares the same fundamental building block as the band-resolved quantum metric. However, the full LPGE response is modulated by the shift vector~\cite{chaudhary2018berry}, which depends on interband phase gradients. As a result, the LPGE response reflects a combination of quantum metric and complex phase-gradient effects, making extraction of the quantum metric challenging~\cite{avdoshkin2025multistate}.

Conventional photogalvanic probes rely on spatially homogeneous 
optical excitation, where crystal symmetry strongly constrains the 
allowed responses: second-order photocurrents are strictly forbidden in 
centrosymmetric systems, while in non-centrosymmetric materials, 
geometric and non-geometric contributions to the response are 
difficult to disentangle. Nonlinear optical responses are inherently 
enhanced near band degeneracies, where interband mixing is maximized 
and quantum geometric effects become prominent. In systems such as 
nodal-line and Weyl semimetals (WSMs) ~\cite{ahn2020low,salerno2020floquet}, quantum geometric 
quantities can become large and dominate the optical response, 
providing an opportunity to probe geometric effects in a controlled 
and tunable manner.

Together, these considerations motivate an optical protocol that can simultaneously isolate Berry curvature and quantum metric channels within the same device while remaining viable in high-symmetry crystals. Such a protocol would apply across a material space not restricted by the symmetry or field requirements of existing routes. Theory~\cite{xie2025photon,ji2024opto} has suggested that nonlinear optical responses with engineered photon momentum can provide cleaner access to quantum geometry, and spatially dispersive CPGE (s-CPGE) has been demonstrated as a probe of Berry curvature--driven responses~\cite{ji2019spatially}, but the complementary quantum metric channel was not isolated within that framework. Here we extend this approach to its full quantum geometric scope using a controllable in-plane optical intensity gradient, resolving momentum-space quantum metric hotspots without requiring special symmetry considerations, while simultaneously detecting both Berry curvature and quantum metric channels from the same device. Because the helicity-even spatially dispersive response is dominated by the injection contribution, whose geometric weight is the quantum metric itself, it provides cleaner access to the metric than the conventional LPGE, where the same information is entangled with shift vector phase-gradient terms (see Supplementary Sec.~S4). This response is also intrinsic in origin: the gradient-induced imbalance in interband injection persists when the relaxation time is uniform across the Brillouin zone, so no scattering asymmetry is required to generate the signal. The geometric weight therefore remains unaffected by disorder and other extrinsic effects.

Here we implement this spatially dispersive photogalvanic effect (sPGE) protocol experimentally and develop the accompanying symmetry-based framework for the resulting geometric photocurrents. Using $1T^\prime$-MoTe$_2$ as a model platform with a temperature-driven structural transition, we track how the response evolves across a well-defined change in inversion symmetry. With polarization-resolved measurements under nonuniform illumination, we separate helicity-even (linear) and helicity-odd (circular) photocurrent spectra to distinguish quantum metric-- and Berry curvature--driven contributions, corresponding to LPGE and CPGE channels, respectively~\cite{wang2026direct}. First-principles calculations are in close agreement with experiments and show that photocurrent extrema coincide with hotspots of the quantum metric (s-LPGE) and Berry curvature (s-CPGE). Finally, via Fermi-level tunability~\cite{rhodes2017engineering}, we demonstrate a systematic blueshift of the photocurrent spectra consistent with a band- and momentum-resolved picture of compensating Brillouin-zone contributions. Our results establish an all-optical route to accessing the QGT in crystalline solids and provide a practical knob to reshape geometric photocurrent spectra toward quantum geometric--enabled optoelectronic functionalities.

\section{Results}

\subsection{Spatially dispersive photocurrent and geometric decomposition}

To probe the connection between spatially dispersive photocurrents and quantum geometry, we investigated the polarization-dependent sPGE response in MoTe$_2$. $1T^\prime$-MoTe$_2$ is an ideal platform for studying geometric photocurrents owing to its temperature-tunable inversion symmetry and strong spin--orbit coupling (SOC), which together drive large asymmetric carrier excitation and relaxation pathways that enhance the photocurrent response~\cite{singh2020engineering}. MoTe$_2$ is a layered material, and each layer possesses mirror ($M_y$) and rotational ($C_2$) symmetries (Fig.~1(a))~\cite{jiang2017signature}. The high-temperature $1T^\prime$ phase is inversion symmetric and suppresses conventional second-order optical effects~\cite{boyd2008nonlinear}. At approximately 250~K, bulk MoTe$_2$ undergoes an interlayer shear distortion (Figs.~1b, 1c)) that breaks inversion symmetry and transitions into the type-II Weyl semimetal $T_d$ phase~\cite{berger2018temperature}, strongly enhancing Berry curvature--driven effects such as CPGE in the mid-infrared regime.

To probe sPGE responses arising from spatial gradients of the optical field, we performed polarization-resolved photocurrent measurements on exfoliated MoTe$_2$ devices at 300~K (inversion-symmetric $1T^\prime$ phase) and 77~K (inversion-broken $T_d$ phase). A focused Gaussian laser beam (full width at half maximum $\sim 30~\mu$m) was scanned across the device (Fig.~1(d)) while its polarization state was continuously modulated using a quarter-wave plate (QWP) and a photoelastic modulator (PEM) (see Methods). In a two-terminal device geometry, the steady-state sPGE current must satisfy
$\nabla \cdot \vb{J} = 0$
due to charge conservation. The detected sPGE signal in the two-terminal geometry is collected by the device electrodes and analyzed following Supplementary Secs.~S1 and~S2F.

At 300~K, the photocurrent exhibits a pronounced dependence on the linear polarization of light while showing nearly identical magnitudes for left- and right-handed circular polarization ($\phi = 45^\circ$ and $135^\circ$), indicating a negligible circular-polarization contribution (Fig.~1(e)). Importantly, when the excitation spot is displaced laterally from position A to position B on the same device, the linear-polarization-dependent photocurrent reverses sign while maintaining a comparable magnitude (Fig.~1(f)). In contrast, the circular-polarization-dependent component remains vanishingly small at both positions. The observed polarization-dependent photocurrent can be fitted phenomenologically by
\[
J = J_C \sin(2\phi) + J_L \sin(4\phi + \delta) + J_0,
\]
where $J_L$ and $J_C$ denote the linear and circular photogalvanic contributions, respectively.

Fits to the experimental data yield opposite signs of $J_L$ at positions A and B, while $J_C \approx 0$ within experimental uncertainty. The sign reversal under lateral displacement demonstrates that s-LPGE is inherently position dependent and driven by the intensity gradient---a hallmark that directly rules out conventional plane-wave photogalvanic mechanisms, for which the response sign is determined solely by crystal symmetry and is position independent~\cite{ji2019spatially}.

Meanwhile, upon lowering the temperature to 77~K, we observe a clear increase in the circular component $J_C$ that emerges sharply as soon as the structural transition occurs from the inversion-symmetric $1T^\prime$ phase to the inversion-broken $T_d$ phase (Fig.~1(g)), while the linear component $J_L$ changes only weakly (see Supplementary Sec.~S10). This behavior signals activation of an intensity-gradient-induced s-CPGE at low temperature. Although conventional CPGE should vanish under normal incidence due to the $C_2$ symmetry in the $a$--$c$ plane and the $M_y$ symmetry in the $a$--$b$ plane of the crystal, the optical intensity gradient activates the s-CPGE response~\cite{ji2019spatially}. The s-CPGE signal exhibits linear scaling with laser intensity (Fig.~1(h)), consistent with the quadratic dependence of a second-order nonlinear optical process on the optical field.

Unlike conventional PGE under uniform illumination, s-LPGE and s-CPGE arise from gradients of the optical field. Under uniform illumination, the quantum geometric contribution to the photocurrent cancels upon Brillouin-zone integration because the current is weighted by the group velocity $v(\mathbf{k})$, which is odd under $\mathbf{k}\rightarrow -\mathbf{k}$. A focused beam supplies finite in-plane photon momentum, which asymmetrically weights interband excitations across momentum space and lifts this cancellation. Opposite crystal momenta no longer contribute equally, producing a photocurrent proportional to the local intensity gradient. Since interband excitation is governed by the quantum geometry of the Bloch states, the resulting injection current directly reflects contributions from the quantum geometric tensor. Because the imbalance resides in the interband excitation rate itself rather than in the relaxation channel, it survives when the relaxation time is uniform across the Brillouin zone.

To leading order in the optical-field gradient $\mathbf{q}$, the sPGE tensor can be decomposed into polarization-resolved channels:
\[
\beta_{lij}^{\mathrm{circular}}
\propto
q_l
\sum_{\mathbf{k},n,m}
f_{nm}(\mathbf{k})
\,
\Omega_{nm}^{ij}(\mathbf{k}),
\]
\[
\beta_{lij}^{\mathrm{linear}}
\propto
q_l
\sum_{\mathbf{k},n,m}
f_{nm}(\mathbf{k})
\,
g_{nm}^{ij}(\mathbf{k}),
\]
where $q_l$ is the optical-field gradient along direction $l$, $f_{nm}(\mathbf{k}) = f_n(\mathbf{k}) - f_m(\mathbf{k})$ is the occupation difference between bands $n$ and $m$, $\Omega_{nm}^{ij}(\mathbf{k})$ is the band-resolved Berry curvature, and $g_{nm}^{ij}(\mathbf{k})$ is the band-resolved quantum metric. Non-geometric prefactors have been suppressed for clarity (see Supplementary Sec.~S2 for the full tensorial expressions). The helicity-odd (circular) response is associated with Berry curvature--related contributions, while the helicity-even (linear) response is dominated by quantum metric--related terms.

Therefore, in the inversion-symmetric $1T^\prime$ phase, combined $PT$ symmetry forces the Berry curvature to vanish identically across the Brillouin zone, suppressing the helicity-odd (circular) channel. The quantum metric, however, is symmetry-even and remains finite, permitting a finite helicity-even s-LPGE response even at room temperature. Upon cooling into the inversion-broken $T_d$ phase, combined $PT$ symmetry is lifted, Berry curvature contributions become finite, and an s-CPGE response is activated. The robust s-LPGE response and the low-temperature emergence of s-CPGE together establish that a controlled optical intensity gradient provides experimental access to both quantum geometric channels.

\subsection{Spectral and momentum-space signatures of the quantum geometric tensor}
The measurements so far probe the two channels only through their symmetry behavior. To test the geometric content of the response functions directly, we measured the photon-energy dependence of the s-LPGE and compared it with first-principles calculations, connecting the measured spectral features to the momentum-resolved quantum metric of the participating bands. By extracting the metric-dominated response component at 300 K (inversion-symmetric 1T$'$ phase) while scanning the excitation wavelength across the mid-infrared range, we observed a reproducible spectral structure (Fig.~2a). The s-LPGE response decreases gradually toward longer wavelengths while exhibiting clear peak features near $\sim 8.6~\mu$m and $10.8~\mu$m. A similar spectral profile is obtained in the inversion-broken $T_d$ phase at 77 K (Fig.~2b), including both the overall decay toward longer wavelengths and the peak structure within the same spectral window. This similarity indicates that the dominant energy scale governing the s-LPGE response is not controlled by inversion symmetry or Berry curvature selection rules. Instead, it points to a structural-phase-robust contribution governed by the quantum metric, which is symmetry-even and therefore remains active through spatial dispersion in both phases.

To understand the electronic origin of this phase robustness, we compare the calculated low-energy band structures of the centrosymmetric 1T$'$ and non-centrosymmetric $T_d$ phases (Fig.~2c). The two phases exhibit quite similar electronic structures overall, consistent with the nearly unchanged s-LPGE spectra across the structural transition~\cite{berger2018temperature}. At the same time, the inversion-breaking distortion in the $T_d$ phase lifts spin degeneracies through spin-orbit coupling and produces Rashba-like band splittings near the Fermi level~\cite{sun2015prediction, wang2016mote2}. This reconstruction generates Weyl points in the low-temperature phase, as predicted by first-principles calculations and confirmed experimentally~\cite{sun2015prediction,deng2016experimental}. Although this symmetry breaking is essential for the emergence of the s-CPGE discussed below, it does not substantially reconstruct the broader interband manifold responsible for the s-LPGE response. This picture is further supported by the momentum-resolved quantum metric near the Weyl-point regions (Fig.~2c, inset). Across the two phases, the overall quantum metric distribution remains similar, consistent with the nearly identical helicity-even photocurrent spectra observed experimentally. In the $T_d$ phase, however, the quantum metric develops a more singular and strongly enhanced character in the immediate vicinity of the Weyl-point region. The structural transition therefore introduces a localized geometric reorganization associated with inversion breaking and Weyl-node formation while leaving the broader quantum metric landscape governing the s-LPGE largely intact.

We next compare the experimental s-LPGE spectra with first-principles calculations of the leading \(q\)-linear, injection-dominated spatially dispersive response, evaluated from Wannier-interpolated band structures in the long-wavelength gradient limit (details in Supplementary Sec.~S5). Despite their distinct symmetries, the two phases yield nearly identical calculated s-LPGE spectra (Fig.~2d), reflecting the close similarity of their momentum-resolved quantum metric. The calculations reproduce both the overall lineshape and the peak structure observed experimentally (Figs.~2a, ~2b,~2d). The calculations contain no scattering asymmetry and treat the relaxation time as uniform across the Brillouin zone; that they nevertheless reproduce the measured lineshape and peak structure supports the intrinsic origin of the s-LPGE. The calculated features appear sharper than experiment because relaxation-time broadening is not included. The dominant discrepancy between theory and experiment is a modest shift of the peak positions by approximately $\sim 2~\mu$m, which we attribute to known limitations of standard density-functional methods in describing low-energy interband separations~\cite{verma2020status}. A direct numerical comparison of the injection and shift contributions shows that the s-LPGE is dominated by the injection currents under the present experimental conditions (Supplementary Sec.~ S5C). The measured response is therefore weighted by the band-resolved quantum metric itself, rather than by the shift vector, whose interband phase gradients enter the conventional LPGE and obscure the metric-related response.

To directly connect the observed spectral peaks to their momentum-space geometric origin~\cite{gao2019nonreciprocal,das2023intrinsic}, we compare the Brillouin-zone hotspots dominating the s-LPGE integrand with the momentum-resolved quantum metric of the valence and conduction bands at the calculated peak energy (Fig.~2e). The spatial patterns closely coincide, demonstrating that the s-LPGE selectively weights regions of enhanced quantum metric~\cite{liu2025quantumgeometry}. This correspondence persists across multiple spectral features (see Supplementary Sec.~S5D). Tuning the photon energy therefore provides energy-selective access to the quantum metric of specific interband transitions, linking the measured spectra to the quantum geometry of the underlying Bloch states.

We now turn to the helicity-odd channel and its Berry curvature origin. In contrast to the s-LPGE, the s-CPGE is observed only in the low-temperature $T_d$ phase, where inversion symmetry is broken and Berry curvature is symmetry-allowed. We therefore focus on measurements at 77 K and examine the s-CPGE response as a function of excitation wavelength (Fig.~3a). The s-CPGE photocurrent decreases with increasing wavelength, reaches a pronounced minimum near $\sim 9 ~\mu$m to $10~\mu$m, and then rises again at longer wavelengths. First-principles calculations of the Berry curvature--driven s-CPGE reproduce this nonmonotonic trend, capturing both the overall decay and the recovery beyond the dip (Fig.~3b), with the characteristic $\sim 2~\mu$m spectral shift consistent with that observed in the s-LPGE calculations.

The pronounced dip near $\sim 9 ~\mu$m to $10~\mu$m in the s-CPGE spectrum --- a feature that also appears in the s-LPGE --- reveals where QGT-driven contributions compete across momentum space. We analyze the calculated s-CPGE at a representative dip energy (blue arrow in Fig.~3b). The corresponding momentum-resolved response (Fig.~3c) shows competing velocity-weighted carrier contributions of opposite sign arising from different regions of the Brillouin zone that nearly cancel in the total integral. In particular, the dominant positive contribution originates from states along the $\Gamma$--Y direction, while a negative contribution emerges along $\Gamma$--X. This compensation mechanism explains why the helicity-odd current becomes strongly suppressed within this wavelength range even though the Berry curvature itself remains finite: the optical excitation selectively probes momentum-space regions with opposite weights and comparable curvature-driven response, leading to near cancellation of the net current.

At the peak energy near $\sim 10.7~\mu$m (black arrow in Fig. 3b), we compare the calculated momentum-resolved s-CPGE with the Berry curvature of the bands participating in the dominant optical transitions. The two maps share the same overall hotspot structure, confirming that the Berry curvature provides the geometric backbone of the helicity-odd response, in direct analogy to the quantum metric correspondence of the s-LPGE channel. Both the s-LPGE and s-CPGE originate from the same set of optically allowed interband transitions, and their distinction emerges primarily through polarization selection: the helicity-even and helicity-odd components of the sPGE project predominantly onto the symmetric (quantum metric) and antisymmetric (Berry curvature) sectors of the quantum geometric tensor, respectively. Consistent with this common microscopic origin, both channels exhibit similar overall spectral trends and share a dip feature within the same mid-infrared wavelength range. This reflects the influence of the same non-geometric weighting factors --- including optical matrix elements, joint density of states, and band velocities --- that enter the momentum-space integration in both channels.

The s-CPGE (helicity-odd channel) is nonetheless substantially more sensitive than the s-LPGE to momentum-space cancellation and interference: contributions of opposite sign from nearby Brillouin-zone regions compete strongly, with their relative weighting set by the detailed structure of interband matrix elements, band velocities, and scattering processes. We further note that although Weyl nodes individually host strongly enhanced quantum metric and Berry curvature and can locally amplify both the s-LPGE and s-CPGE responses, they necessarily appear in pairs with opposite chirality whose contributions largely cancel upon Brillouin-zone integration. This cancellation is a general feature of both channels, such that the net photocurrent is governed primarily by the broader momentum-space structure of the active interband manifold rather than by local Weyl-node enhancement alone (Supplementary Sec.~S5E). The emergence of the s-CPGE in the $T_d$ phase instead reflects the global activation of Berry curvature upon inversion-symmetry breaking, while the s-LPGE remains large across both phases. Both channels are further amplified by the strong spin-orbit coupling in MoTe$_2$, which enhances the Berry curvature and quantum metric across the active interband manifold.

\subsection{Doping-induced tuning of geometric photocurrents}

Since s-CPGE is sensitive to interference and momentum space cancellation effects, we next test whether the curvature-driven s-CPGE channel can be tuned through modification of the electronic filling. Previous ARPES studies on Mo$_{1-x}$W$_x$Te$_2$ have shown that increasing W substitution reduces the valence-conduction band overlap and shifts the conduction-band minimum toward the Fermi level, thereby reshaping the low-energy semimetallic pockets~\cite{jin2018phase}. Because Mo and W are nominally isovalent, this trend should not be interpreted as conventional aliovalent charge doping. Rather, W substitution modifies the metal-$d$-Te-$p$ hybridization, SOC, and structural energetics, producing an effective hole-like reweighting of the near-$E_F$ electronic states. We therefore model the dominant low-energy effect of W substitution by a rigid chemical-potential shift in the first-principles calculations. At 77 K, both 2 \% and 9 \% W-doped MoTe$_2$ exhibit the same overall nonmonotonic spectral dependence observed in undoped MoTe$_2$, while simultaneously displaying a systematic blue shift of the s-CPGE spectrum with increasing W concentration (Figs.~4a,~4b). The characteristic dip feature, which is more clearly resolved in the calculated spectra, shifts toward shorter wavelengths as the W concentration increases, indicating that the helicity-dependent response is highly sensitive to modifications of the low-energy electronic structure.

Motivated by this observation, we model the effect of W doping within first-principles calculations by introducing hole doping and recomputing the Berry curvature--driven s-CPGE spectra. The calculated spectra for two representative filling levels reproduce both the experimentally observed spectral evolution and the blue shift of the dip feature (Figs.~4c,~4d), demonstrating that the evolution of the s-CPGE is governed primarily by doping-induced changes in the electronic filling rather than by extrinsic effects.

To uncover the microscopic origin of the blue shift, we analyze how the momentum-resolved contributions to the s-CPGE reorganize with doping. The blue arrows in Figs.~4c,~4d indicate the dip energies in the calculated spectra, and the corresponding momentum-resolved decompositions evaluated at those energies are shown in Figs.~4e,~4f. In the lower-doping case, the dip originates from near cancellation between opposite-signed velocity-weighted carrier contributions arising predominantly from distinct regions of the Brillouin zone. Upon shifting the Fermi level, however, these competing channels reorganize asymmetrically. The dominant hotspot pattern associated with the Weyl-adjacent contribution remains comparatively robust under moderate filling changes, whereas the competing contribution is strongly reshaped because part of the optically active manifold is removed from the excitation window as the chemical potential shifts. Consequently, the near-cancellation condition no longer occurs at the same photon energy. Instead, compensating the robust contribution requires accessing a broader and higher-energy manifold, thereby shifting the dip toward higher photon energies and producing the experimentally observed blue shift of the s-CPGE response.

This momentum-space picture identifies the dip position as a sensitive spectroscopic marker of competing Berry curvature contributions across the Brillouin zone. Because the Berry curvature changes sign across the Brillouin zone while the quantum metric is symmetry-even, the curvature-driven channel is intrinsically subject to cancellation in a way the metric-driven channel is not. At the dip, contributions from different momentum-space regions enter with opposite velocity weighting and comparable magnitude, so the net current nearly vanishes; shifting the Fermi level redistributes these contributions asymmetrically, moving the cancellation condition to a different photon energy and making the dip position highly sensitive to electronic filling. Tuning the electronic filling therefore provides direct control over how Weyl-proximate Berry curvature contributions are weighted within the nonlinear optical response. In contrast, the metric-dominated channel does not exhibit comparable doping evolution at room temperature: the s-LPGE spectrum remains nearly unchanged within experimental resolution across the same doping series (Supplementary Sec.~S9).

The effect of W substitution is thus a rigid shift of the excitation window, and the quantitative agreement between the calculated and measured shift of the dip confirms that this feature originates from the Berry-curvature-weighted interband response.

\section{CONCLUSION}
In summary, we have demonstrated that the spatially dispersive photogalvanic effect provides direct access to both the quantum metric and Berry curvature within a single device, without requiring modification of the underlying crystal symmetries, applied fields, or magnetic order. The response is intrinsic: it persists even when the relaxation time is taken uniform across the Brillouin zone. The close agreement between experiment and first-principles calculations, together with the demonstrated tunability through W doping, establishes spatially varying optical fields as a practical and broadly applicable framework for probing the quantum metric and Berry curvature hotspots in momentum space. The quantum metric, which has no clean linear-polarization photocurrent channel under uniform illumination, thus becomes directly accessible in the bulk. More generally, the spectral structure of the spatially dispersive photocurrent provides an energy-resolved window into the quantum geometry of Bloch states, opening a pathway toward selectively accessing Berry curvature and quantum metric contributions in a wide range of materials, including topological semimetals, kagome systems with flat bands, and moiré superlattices.

\section*{Acknowledgments}
The initial phase of the work was supported by the US Air Force Office of Scientific Research (award No. FA9550-20-1-0345) followed by support from the Office of Naval Research (grant No. N000142512360) and the National Science Foundation (grant No. DMR-2508192 and via the seed grant from DMR-2309043). Device fabrication and characterization work were carried out at the Singh Center for Nanotechnology, which is supported by the NSF National Nanotechnology Coordinated Infrastructure Program under grant NNCI-1542153. B.K. and A.M.R. acknowledge support from the U.S. Department of Energy, Office of Science, Basic Energy Sciences, under Award No. DE-SC0024942. 

\bibliography{reference}

\newpage

\begin{figure*}
    \centering
    \includegraphics{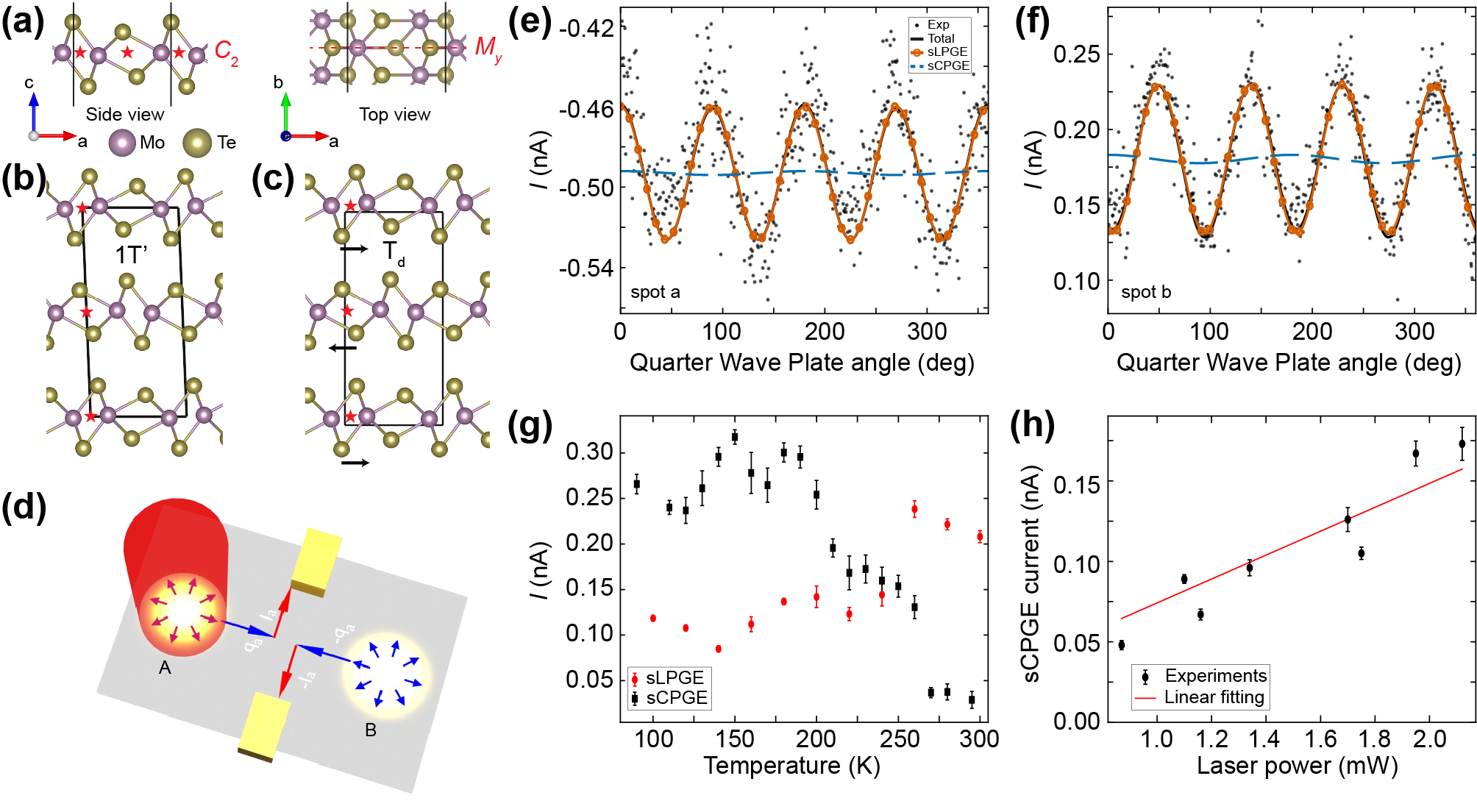}
    \caption{
Spatially dispersive photogalvanic responses in MoTe$_2$. (a) Side and top views of a single-layer MoTe$_2$ crystal. The twofold rotational axis and mirror plane are denoted as $C_2$ and $M_y$, respectively. (b,c) Crystal structures of the centrosymmetric 1T$'$ phase (room temperature) and the non-centrosymmetric $T_d$ (Weyl) phase (below 250 K). (d) Schematic of the spatially dispersive photocurrent measurement geometry. The red cylinder denotes the focused mid-infrared beam, the blue arrow indicates the in-plane intensity gradient q, and the red arrow the resulting photocurrent J. Displacing the beam laterally from position A to position B reverses the averaged intensity gradient and correspondingly reverses the direction of current. (e,f) Room-temperature polarization-dependent photocurrent measured at two spatial positions, A (e) and B (f), under focused Gaussian-beam illumination. The total sPGE response is decomposed into helicity-even (s-LPGE) and helicity-odd (s-CPGE) components. The linear (s-LPGE) contribution reverses sign upon lateral displacement of the excitation spot, while the circular (s-CPGE) component remains negligible. (g) Temperature dependence of the s-LPGE and s-CPGE responses across the structural phase transition. (h) Laser-power dependence of the s-CPGE measured at a wavelength of $7.1~\mu$m, showing a linear scaling with excitation power.
}
    \label{fig1}
\end{figure*}

\newpage

\begin{figure*}
    \centering
    \includegraphics{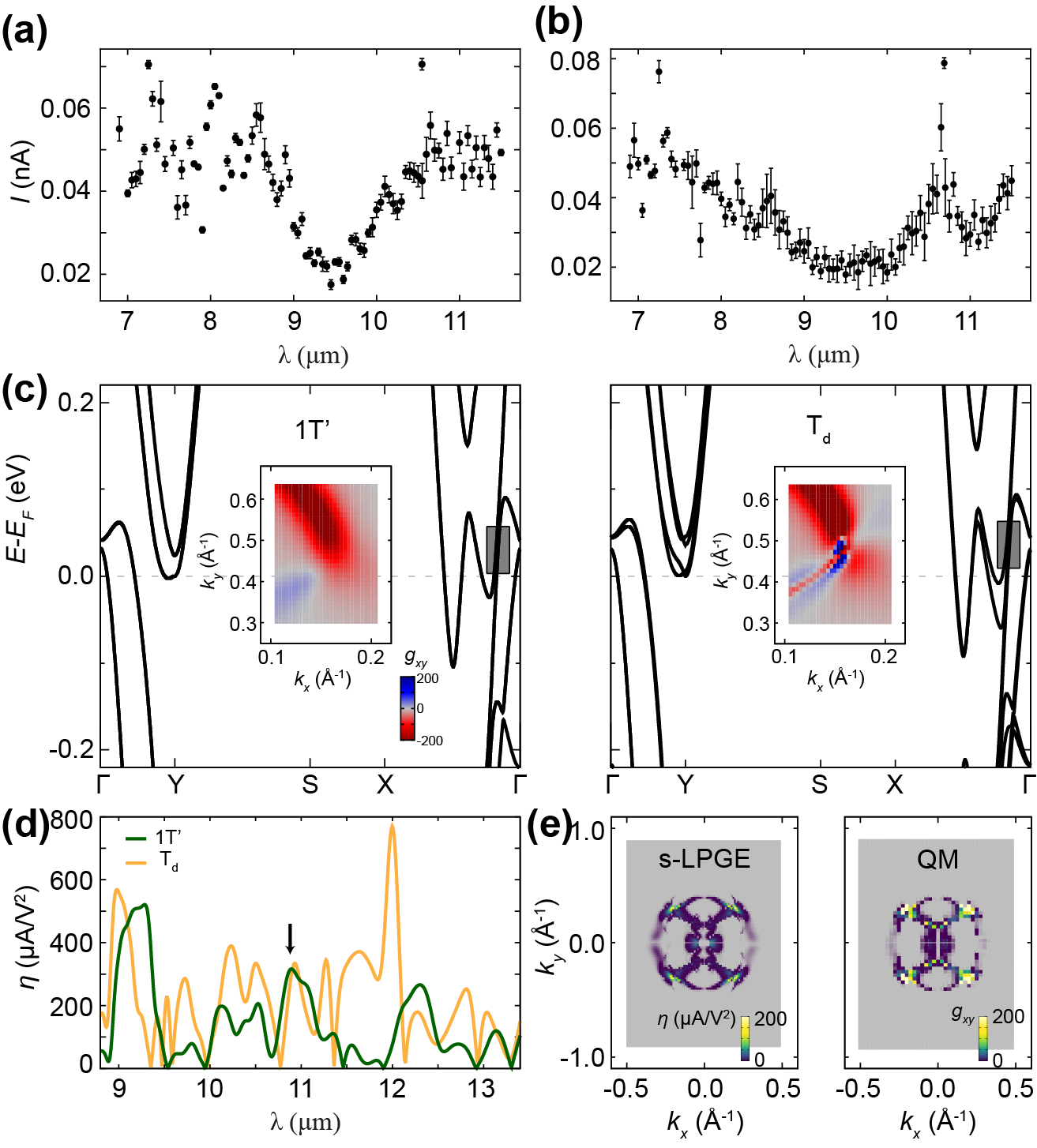}
    \caption{
Spectral and momentum-space signatures of quantum metric--driven photocurrent.
(a,b) Wavelength dependence of the spatially dispersive LPGE (s-LPGE) measured at 300 K(a) and 77 K(b), respectively, showing closely similar line shapes and characteristic spectral features across the structural transition.
(c) Calculated electronic band structure in the relevant energy window, indicating the interband transitions contributing to the measured response. Inset: Momentum-resolved quantum metric near the Weyl points for the two phases.
(d) First-principles calculations of the s-LPGE spectra for the $1T'$ and $\mathrm{T_d}$ phases, reproducing the experimental line shape and peak structure.
(e) Comparison between the momentum-resolved s-LPGE response and the band-resolved quantum metric at the peak energy, showing closely matching hotspot distributions.
}
    \label{fig2}
\end{figure*}

\newpage

\begin{figure*}
    \centering
    \includegraphics{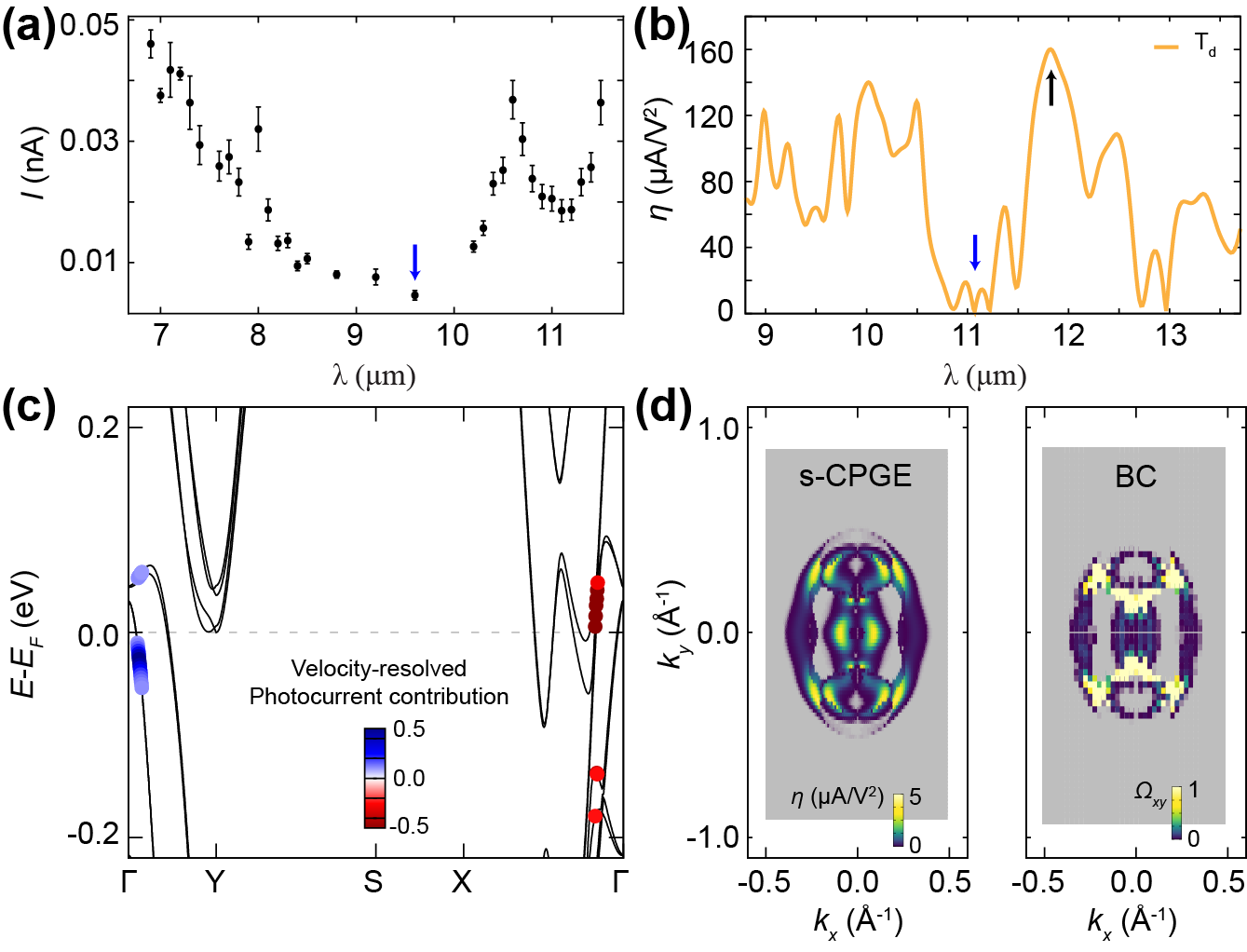}
    \caption{
Berry curvature--driven s-CPGE response and competing momentum-space contributions. (a) Wavelength dependence of the spatially dispersive CPGE (s-CPGE) measured at 77 K, showing a nonmonotonic spectral response with a pronounced minimum near $\sim 9 ~\mu$m to $10~\mu$m. (b) First-principles calculation of the s-CPGE spectrum, reproducing the overall nonmonotonic spectral evolution. The blue and black arrows indicate the dip and peak energies analyzed in (c) and (d), respectively. (c) Velocity-weighted photocarrier contributions to the s-CPGE at the dip energy across the Brillouin zone, showing positive (red) and negative (blue) contributions from different momentum-space regions that nearly cancel in the total response due to opposite-sign weighting factors. (d) Comparison between the momentum-resolved s-CPGE response and the Berry curvature of the relevant bands participating in the dominant optical transitions, showing strong similarity.
}
    \label{fig3}
\end{figure*}

\newpage

\begin{figure*}
    \centering
    \includegraphics{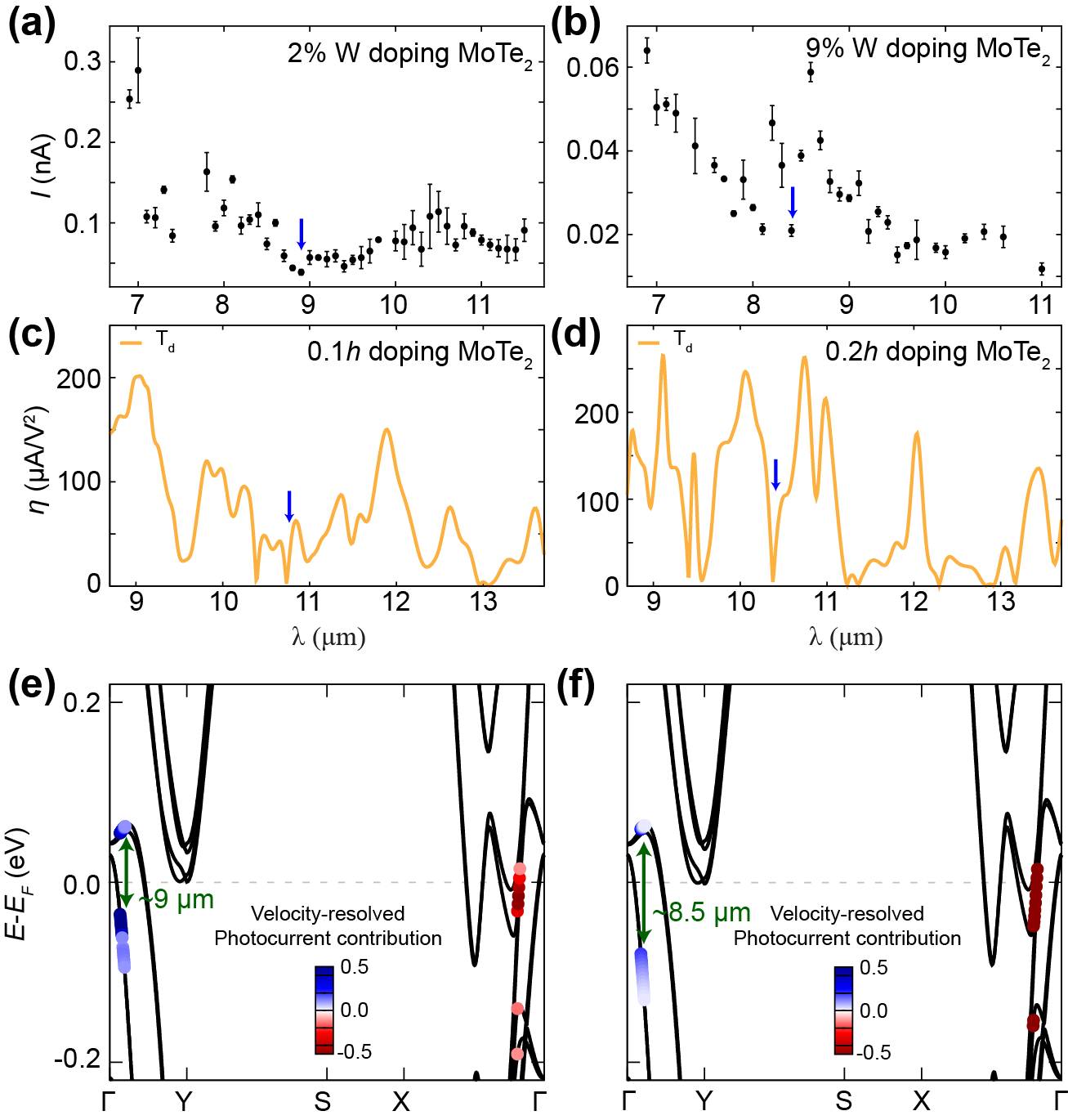}
    \caption{
Doping dependence of the Berry curvature--driven s-CPGE response. (a,b) Wavelength dependence of the spatially dispersive CPGE (s-CPGE) measured at 77 K for 2 \% (a) and 9 \% (b) W-doped MoTe$_2$, showing a systematic shift of the characteristic spectral dip (blue arrows), which is more clearly resolved in the calculated spectra shown in (c,d). (c,d) Corresponding first-principles calculations reproducing the experimentally observed spectral evolution and the doping-induced shift of the dip position with increasing hole doping. (e,f) Velocity-weighted photocarrier contributions to the s-CPGE evaluated at the dip energies for $0.1~h$ (e) and $0.2~h$ (f) hole doping, where the chemical potential is rigidly shifted in units of holes. The calculations reveal an asymmetric redistribution of momentum-space contributions that shifts the cancellation condition toward higher photon energies with increasing doping.
}
    \label{fig4}
\end{figure*}

\newpage

\end{document}